\documentclass[
	aps,
	prstab,
	a4paper,
	amsmath,
	amssymb,
	floatfix,
	showpacs,
	reprint
	]{revtex4-1}

\usepackage[utf8]{inputenc}
\usepackage{graphicx}
\usepackage{siunitx}
\usepackage{bm}
\usepackage{subcaption}
\usepackage{hyperref}

\graphicspath{{./graphics/}{./}}

\newcommand*{\AFFUNI}{University of Hamburg, Department of Physics, Luruper Chaussee 149, 22761 Hamburg, Germany}
\newcommand*{\AFFDESY}{Deutsches Elektronen-Synchrotron DESY, Notkestraße 85, 22603 Hamburg, Germany}
\renewcommand*{\vec}[1]{\bm{#1}}
\newcommand*{\deriv}[2]{\dfrac{\textrm{d}#1}{\textrm{d}#2}}
\newcommand*{\pderiv}[2]{\dfrac{\partial#1}{\partial#2}}
\newcommand*{\mat}[1]{\mathcal{#1}}
\newcommand*{\figref}[1]{\figurename~\ref{#1}}
\newcommand*{\refcite}[1]{Ref.\@ \cite{#1}}

\begin{document}

\title{Confining continuous manipulations of accelerator beamline optics}

\author{Ph.\@ Amstutz}
\email{philipp.amstutz@desy.de}
\author{C. Lechner}
\author{T. Plath}
\affiliation{\AFFUNI}

\author{S. Ackermann}
\author{J. Bödewadt}
\author{M. Vogt}
\affiliation{\AFFDESY}

\date{\today}

\begin{abstract}
Altering the optics in one section of a linear accelerator beamline will in general cause
an alteration of the optics in all downstream sections.
In circular accelerators, changing the optical properties of any beamline element
will have an impact on the optical functions throughout the whole machine.
In many cases, however, it is desirable to change the optics in a certain beamline section
without disturbing any other parts of the machine.
Such a local optics manipulation can be achieved by adjusting a number of additional corrector magnets
that restore the initial optics after the manipulated section.
In that case, the effect of the manipulation is confined in the region
between the manipulated and the correcting beamline elements.
Introducing a manipulation continuously, while the machine is operating,
therefore requires continuous correction functions to be applied to the correcting quadrupole magnets.
In this paper we present an analytic approach to calculate such continuous correction functions for six
quadrupole magnets by means of a homotopy method.
Besides a detailed derivation of the method, we present its application to an algebraic example,
as well as its implementation at the seeding experiment sFLASH at the free-electron laser FLASH located at DESY in Hamburg.
\end{abstract}


\maketitle

\section{Introduction}
Many applications of particle accelerators require the dynamic manipulation of the
optical functions in certain regions of the beamline during operation
as, for example, minimizing the $\beta$-function at the interaction point of a collider experiment,
or closing a variable-gap undulator in a synchrotron light source or a free-electron laser (FEL). 
Changing the optical properties of a beamline element, however, causes not only
local changes in the optics but has an impact on the optical functions in all
sections downstream of the adjusted element.
Hence, in many cases a manipulation will result in unmatched optics in these sections and a
correction is required to rematch the optics.

Our approach is to completely confine the influence of the manipulation to
a region around the adjusted elements, making the manipulation transparent for
all downstream sections.
This can be achieved by appropriately adjusting adjacent quadrupole magnets.
Accelerator simulation codes such as \texttt{ELEGANT}~\cite{Borland2000} or \texttt{MAD-X}~\cite{MADX} can perform this matching
and determine a suitable correction by numerically solving minimization problems.
Such an approach, however, is ill-suited for calculating continuous correction functions,
which are required to compensate a manipulation that is being introduced continuously, as we will discuss later.
In contrast to this, we present an analytic approach that is based on the implicit function theorem
(compare for instance \cite{Krantz2002}) and allows for the determination of correction
parameters as a continuous function of the introduced disturbance.

\begin{figure}[bth]
    \centering
    \includegraphics*[width=\linewidth]{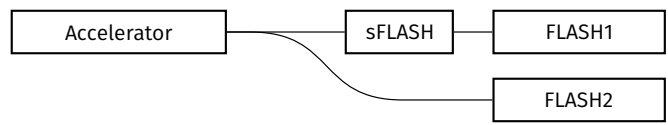}
    \caption{\label{fig:flash}Schematic layout of the FLASH facility.
    The superconducting linear accelerator delivers electron bunches for the FLASH1 and FLASH2
    undulator beamlines and for the seeding experiment sFLASH.}
\end{figure}

Moreover, the implementation of this correction method at the FEL user facility FLASH is presented.
As depicted in \figref{fig:flash}, FLASH features two parallel undulator beamlines FLASH1 and FLASH2.
Upstream of the FLASH1 main undulator the sFLASH experiment is located,
which, since its installation in 2010, utilizes a variable-gap undulator system to generate
seeded FEL radiation using different seeding techniques~\cite{Boedewadt2015}.
By closing this undulator in order to start seeded operation,
the optics in the following FLASH1 main undulator is altered,
which results in a deteriorated FEL performance.
The presented method was developed to provide the means to correct the influence
of the variable-gap undulator and thus allow simultaneous operation of FLASH1 and sFLASH.
Recently, the precise restoration of the initial optics
could be verified by beam size measurements along the FLASH1 main undulator,
which are presented in the final section of this paper.

\section{The method}
In linear approximation the effect of a magnetic structure on the transversal phase space vector
of a particle can be described by a $4\times4$ matrix $\mat{M}$, called the
transport matrix \cite{Courant1958}.
It becomes apparent that any transport matrix is bound to be symplectic, $\mat{M} \in \mathrm{Sp}(4,\mathbb{R})$.
Furthermore it can be shown that the group of symplectic $2N \times 2N$ matrices is a $N(2N+1)$-dimensional manifold~\cite{Gilmore2005},
\begin{equation}
\mathrm{dim}\,\mathrm{Sp}(2N,\mathbb{R}) = N(2N+1),
\label{eq:DOF}
\end{equation}
with $N \in \mathbb{N}$.
As a result of this, any symplectic matrix can be uniquely defined by $N(2N+1)$ of their elements.
A general transversal transfer matrix features therefore $10$ independent elements.
However, in many practical cases the two transversal directions are decoupled.
Then $\mat{M}$ reduces to two symplectic block matrices,
\[ \mat{M} \in \mathrm{Sp}(2,\mathbb{R}) \oplus \mathrm{Sp}(2,\mathbb{R}), \]
each of which consists of 3 independent elements.
Decoupled transfer matrices hence can be fully defined by a total of $6$ elements.

Every beamline can be understood as a sequence of $P$ elements, each of which is represented by a
transport matrix $M_\textrm{n}$. In general, the optical properties of an element
can be altered during operation, e.g. by changing the current of a quadrupole magnet.
Formally, we will therefore treat every transport matrix as a
function of a real experimental parameter $\rho_n$
\[ \mat{M}_\textrm{n} = \mat{M}_\textrm{n}(\rho_n). \]
The total optical properties of a beamline are therefore given by the product
of its constituent's transport matrices
\begin{equation}
\mat{M} = \prod_{n=0}^{P-1} \mat{M}_{P-n}(\rho_{P-n}).
\label{eq:matmult}
\end{equation}
A variation of a subset of $S$ machine parameters
\begin{equation}
\vec{\sigma} = \vec{\sigma}(t) = \left(\sigma_1(t),\dots,\sigma_S(t)\right)^T,
\end{equation}
will therefore result in deviations from the initial optical properties of the beamline.
Here, $\vec{\sigma}(t)$ denotes the course of the disturbance parameters as a function of
a single real parameter $t$.
In many experimental situations such a disturbance is undesirable and the ability
to freely change certain machine parameters without influencing the overall optical properties
of a beamline section would be of great benefit.
To achieve this, the disturbance has to be corrected by another subset of machine parameters $\vec{\tau}$.
These parameters will have to be controlled such that they fulfill the condition
\begin{equation} 
\mat{M}(\vec{\sigma}(t),\vec{\tau}(t)) = \mat{M}(\vec{\sigma}^0,\vec{\tau}^0),
\label{Mconst}
\end{equation} 
where $\vec{\sigma}^0 = \vec{\sigma}(0)$ and $\vec{\tau}^0=\vec{\tau}(0)$ denote the constant initial machine parameters.
If this condition is fulfilled, the transport matrix of the beamline is constant and in particular is not affected by the
disturbances.

As we have seen, $\mat{M}$ is a symplectic matrix and therefore has a reduced number of
independent elements.
By defining a vector-valued function $\vec{B}\colon\,\mathbb{R}^{P}\to\,\mathbb{R}^{N(2N+1)}$ with the components
\[ B_i(\vec{\sigma},\vec{\tau}) = \mat{M}_i(\vec{\sigma},\vec{\tau}) - \mat{M}_i(\vec{\sigma^0},\vec{\tau^0}),\]
where the $\mat{M}_i$ are a set of independent elements of $\mat{M}$,
equation \eqref{Mconst} can be written as
\begin{equation} 
\vec{B}(\vec{\sigma}(t),\vec{\tau}(t)) = 0.
\label{Bconst}
\end{equation} 

Even for simple beamlines sections, consisting only of a few elements, $\vec{B}(\vec{\sigma},\vec{\tau})$ will be highly
intricate function. In general, it is therefore not  possible to solve equation \eqref{Bconst} for $\vec{\tau}(t)$
by algebraic means, given an arbitrary disturbance function $\vec{\sigma}(t)$.
For a given set of numerical values for the disturbance parameters it is, however, possible to find a suitable
approximation for correction parameters by utilizing numerical minimization methods.
This approach is readily realized with the help of existing accelerator simulation codes.
To find a numerical approximation to the correction function using this method, one would have to compute
the correction parameters for a number of disturbance values along $\vec{\sigma}(t)$ and interpolate.
However, two successive interpolation points are uncorrelated, as they were found by two independent numerical processes. 
Hence, even with a high number of interpolation points one can not expect that this method will produce a smooth, well-behaved
correction function.
In contrast to this, we will derive a method that yields continuous correction functions.
Loosely speaking, our approach is to track how a known root of $\vec{B}$ evolves along $\vec{\sigma}(t)$,
rather than to recalculate the root for each step.

The mathematical foundation of this method is the implicit function theorem \refcite{Krantz2002}.
It states, in short, that equation \eqref{Bconst} implicitly defines a unique correction function $\vec{\tau}(t)$ in a neighborhood
around those machine states $(\vec{\sigma}^\star,\vec{\tau}^\star)$, at which the Jacobian matrix is non-singular,
\begin{equation}
\mathrm{det}\,\pderiv{\vec{B}}{\vec{\tau}}(\vec{\sigma}^\star,\vec{\tau}^\star) \neq 0,
\label{eq:JacNZ}
\end{equation}
and equation \eqref{Bconst} is fulfilled
\begin{equation} 
\vec{B}(\vec{\sigma}^\star,\vec{\tau}^\star) = 0.
\end{equation} 
For the implicit function theorem to be applicable $\vec{B}$ and $\vec{\tau}$ have to be of the
same dimension. Only then the Jacobian matrix in equation~\eqref{eq:JacNZ} is square and its determinant can be formed.
The number of disturbance parameters $\vec{\sigma}$, in contrast, is not restricted.
In the case of uncoupled matrices, this means that a disturbance in any number of parameters may be corrected by
adjusting only six correction parameters.

Let us for the moment assume that for a certain $\vec{\sigma}(t)$ the corresponding correction function $\vec{\tau}(t)$ exists.
By differentiating equation \eqref{Bconst} with respect to $t$
\begin{equation}
\deriv{B}{t} = 0 = \pderiv{\vec{B}}{\vec{\sigma}} \deriv{\vec{\sigma}}{t} + \pderiv{\vec{B}}{\vec{\tau}} \deriv{\vec{\tau}}{t},
\end{equation}
an expression for the derivative of the correction function can be obtained
\begin{equation}
\deriv{\vec{\tau}}{t} = -\left(\pderiv{\vec{B}}{\vec{\tau}}\right)^{-1} \pderiv{\vec{B}}{\vec{\sigma}} \, \deriv{\vec{\sigma}}{t}.
\label{eq:diffeq}
\end{equation}

This defines a system of coupled non-linear ordinary differential equations for the components of $\vec{\tau}(t)$.
We have therefore shifted the problem of algebraically solving the intricate system \eqref{Bconst} for $\vec{\tau}(t)$ to
finding the solution of the even more intricate differential equations \eqref{eq:diffeq}.
While it still will be unfeasible to find an analytic solution, this problem now lends itself readily to well-known numerical
approximation methods.
This approach can be considered a numerical homotopy method,
which are extensively treated in \refcite{Allgower2003}.

However, any iterative solution method will inevitably fail at those points where the Jacobian \eqref{eq:JacNZ} is singular.
If the correction function for a certain set of correction parameters contains such a singularity,
these parameters are not suitable to fully compensate the disturbance.
In this case, another set of correction parameters has to be chosen.

\section{An example}
\newcommand{\exB}{
\left(
\begin{array}{c}
 l \left(f_2 \left(f_3 l+2\right)+f_1 \left(2 f_3 l+f_2 l \left(f_3 l+2\right)+3\right)+f_3\right)+1 \\
 l \left(2 f_3 l+f_2 l \left(f_3 l+2\right)+3\right) \\
 3 f_4 l+2 f_3 l \left(f_4 l+1\right)+f_2 l \left(f_3 l+f_4 l \left(f_3 l+2\right)+1\right)+1 \\
\end{array}
\right)
}

\newcommand{\exCOR}{
\left(
\begin{array}{c}
 -\dfrac{f_3 l+2}{\left(f_2 l+2\right) \left(2 f_3 l+f_2 l \left(f_3 l+2\right)+3\right)} \\
 -\dfrac{f_3 l+2}{f_2 l+2} \\
 \dfrac{1}{2 f_3 l+f_2 l \left(f_3 l+2\right)+3} \\
\end{array}
\right)
}

\newcommand{\exSOLa}{\dfrac{f^*_1 \left(f_2 l+2\right) \left(2 f^*_3 l+f^*_2 l \left(f^*_3 l+2\right)+3\right)-\left(f_2-f^*_2\right) \left(f^*_3 l+2\right)}{\left(f_2 l+2\right) \left(2 f^*_3 l+f^*_2 l \left(f^*_3 l+2\right)+3\right)}}
\newcommand{\exSOLb}{\dfrac{f^*_2 \left(f^*_3 l+2\right)+2 f^*_3-2 f_2}{f_2 l+2}}
\newcommand{\exSOLc}{\dfrac{f_2-f^*_2}{2 f^*_3 l+f^*_2 l \left(f^*_3 l+2\right)+3}+f^*_4}

\begin{figure}[!b]
    \centering
    \begin{subfigure}[t]{\linewidth}
        \centering
        \includegraphics[width=0.9\textwidth]{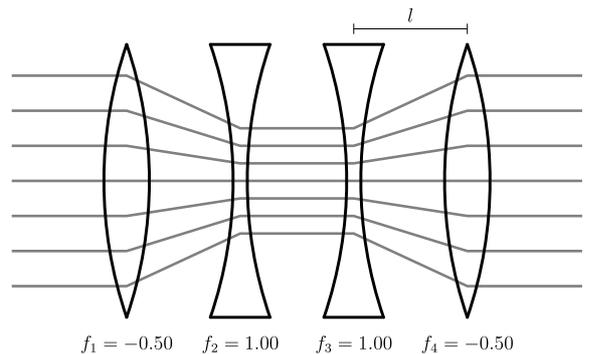}
        \caption{\label{fig:TLXa}Exemplary initial configuration.}
    \end{subfigure}
    \begin{subfigure}[t]{\linewidth}
        \centering
        \includegraphics[width=0.9\textwidth]{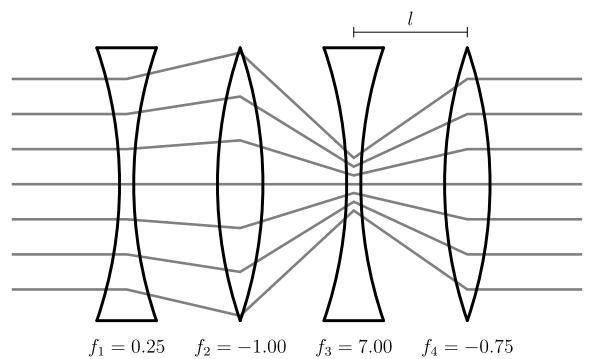}
        \caption{\label{fig:TLXb}Corrected state. Despite the change of sign of $f_2$, the overall optical properties are identical to the initial state.}
    \end{subfigure}
    \caption{Schematic representation of the four-lens system in different configurations.}
\end{figure}
Symbolic transfer matrices of real-world lattices typically consist of long analytical expressions,
which renders obtaining solutions to \eqref{eq:diffeq} very challenging.
To promote better understanding of the proposed method we shall nevertheless
provide the reader with an algebraic, worked out example of application.
For this purpose, we will apply our method to a beamline consisting of four quadrupole magnets, each separated
by a distance $l$. To simplify the calculations in this example, we restrict the problem
to preserving the optical properties in just one plane of motion and treat the quadrupole magnets as thin lenses.
The symbolic $2\times 2$ transfer matrix $\mat{M}$ in this plane of motion is given by
\[ \mat{M} = \mat{L}(f_4)\cdot \mat{D}(l)\cdot \mat{L}(f_3)\cdot \mat{D}(l)\cdot \mat{L}(f_2)\cdot \mat{D}(l)\cdot \mat{L}(f_1), \]
where $\mat{L}(f)$ is the known transport matrix of a thin lens with focal strength $f$,
while $\mat{D}(l)$ represents a drift space,
\[ \mat{L}(f) = \begin{pmatrix} 1 & 0 \\ f & 1 \end{pmatrix} \qquad \mat{D}(l) = \begin{pmatrix} 1 & l \\ 0 & 1 \end{pmatrix}.\]

Suppose the system is in an arbitrary initial configuration $f_i = f_i^*$, as for example depicted in \figref{fig:TLXa}.
If now the focal strength of one lens is altered, the resulting optics will deviate from the initial case.
The goal of this example is to calculate functions for the focal strengths $f_1$, $f_3$ and $f_4$
that correct any influence variation of $f_2$ has on the transfer matrix of the system.

According to equation \eqref{eq:DOF}, $\mat{M}$ has three independent elements.
Therefore, we can choose three arbitrary elements of $\mat{M}$ for the definition of $\vec{B}$.
We might choose
\begin{gather*}
\vec{B} = (\mat{M}_{11},\mat{M}_{12},\mat{M}_{22})^T = \\
\scalebox{0.95}{$\exB$}.
\end{gather*}

We define the correction parameters to be the focal strengths of the remaining lenses
\[ \tau = (f_1,f_3,f_4)^T. \]

In this case, equation \eqref{eq:diffeq} yields the following differential equations
{
\renewcommand*{\arraystretch}{2.25}
\[
\begin{pmatrix}
\deriv{f_1}{f_2}\\
\deriv{f_3}{f_2}\\
\deriv{f_4}{f_2}
\end{pmatrix}
=
\exCOR.
\]
}
The solution to these equations for the initial conditions $f_i(f_2^*) = f_i^*$ are
{
\footnotesize
\begin{align*}
f_1 &= \exSOLa\\
f_3 &= \exSOLb\\
f_4 &= \exSOLc.
\end{align*}
}%

For every value of $f_2$ these functions yield values for the remaing focal strengths,
so that the optical properties of the system are equivalent to those of the initial setting $f_i^*$,
see for example \figref{fig:TLXb}.
It can be seen  that the first two functions have a pole at $f_2 = -2/l$,
independent on the initial condition.
At this point the Jacobian determinant in equation \eqref{eq:JacNZ} vanishes
and no continuous correction beyond this point is possible.

\section{Implementation at \lowercase{s}FLASH}
\begin{figure*}[bth]
    \centering
    \includegraphics*[width=\linewidth]{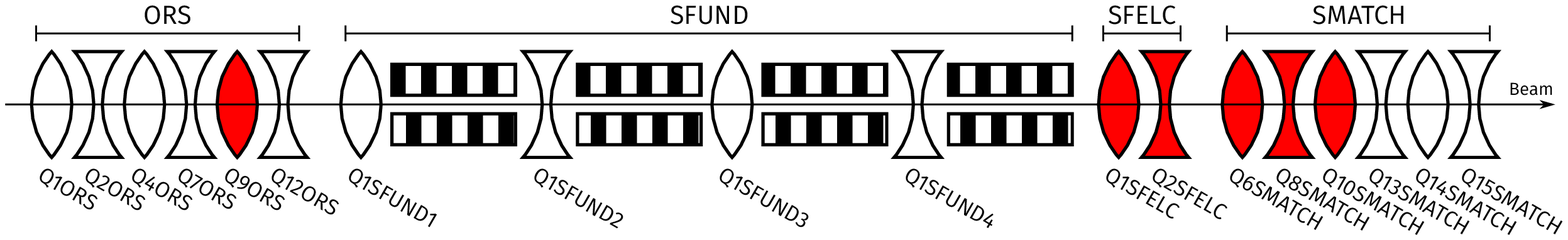}
    \caption{\label{fig:magnets}Positions of the quadrupole magnets close to the SFLASH undulator.
    Note that the magnets in the SFUND section can not be used as correctors.
    Magnets used in the example shown in \figref{fig:correction} are highlighted in red.
    Drawing not to scale.
    }
\end{figure*}

The super-conducting linear accelerator FLASH at DESY in
Hamburg delivers trains of electron bunches into two parallel undulator beamlines, FLASH1 and FLASH2.
While FLASH2 is currently being prepared for user operation,
FLASH1 has been delivering soft x-ray photons for user experiments
since 2005\,\cite{Honkavaara2014}.
Upstream of the FLASH1 main
undulator an experimental setup for FEL seeding (sFLASH) is located, see \figref{fig:flash}.
It features four
variable-gap undulator modules and thus is the ideal test bench for the study of
several different seeding schemes \cite{Boedewadt2015}.

Closing the seeding undulator intensifies its natural focusing effect
and will therefore impact the beam optics further downstream of this undulator
and in particular at the entrance of the main undulator.
At lower beam energies the change in the $\beta$-function can easily be in the order of several tens of
percent so that the matching into the FODO structure of the main undulator is corrupted.
The result is a significantly deteriorated FEL performance.
Therefore, the disturbance introduced by the seeding undulator has
to be compensated by adjusting the field strength of appropriate quadrupole magnets.

Initially, the presented  method was developed to correct for
the influence of the variable-gap undulator used by the seeding experiment at FLASH.
Due to their magnetic structure, planar undulators exhibit a systematic focusing effect
in the vertical plane (here, denoted by $y$), perpendicular to their pole faces.
For our purpose, the additional weak defocusing in the horizontal direction ($x$),
stemming from the finite width of real-world undulators, can be neglected.
Employing the matrix formalism introduced above, the effect of an undulator of length $l$ and
undulator period $\lambda_\mathrm{u}$ on a particle's trajectory vector
$u = \left(x,x',y,y'\right)^T$ is represented by the transfer matrix
\begin{equation}
\mat{M}_{\mathrm{und}} =
\begin{pmatrix}
1 & l & 0 & 0 \\
0 & 1 & 0 & 0 \\
0 & 0 & \mathrm{cos}\sqrt{\kappa}l & \frac{\mathrm{sin}\sqrt{\kappa}l}{\sqrt{\kappa}} \\
0 & 0 & -\sqrt{\kappa}\,\mathrm{sin}\sqrt{\kappa}l & \mathrm{cos}\sqrt{\kappa}l 
\end{pmatrix},
\end{equation}
where $\kappa$ is the focusing strength
\begin{equation}
\kappa \approx 2\left(\dfrac{K\,\pi}{\gamma\lambda_\mathrm{u}}\right)^2,
\end{equation}
defined by the undulator parameter $K$ and $\gamma$, the particle's Lorentz factor \cite{Quattromini2012}.
By varying the value of the gap, the $K$ parameter of the undulator system used by the sFLASH experiment
can be set to a value between $0$ and $2.72$ \cite{Tischer2010}, introducing the undesirable focusing effect.
We therefore identify $K$ as the disturbance parameter.

\begin{figure}[bth]
    \centering
\includegraphics*[width=\linewidth]{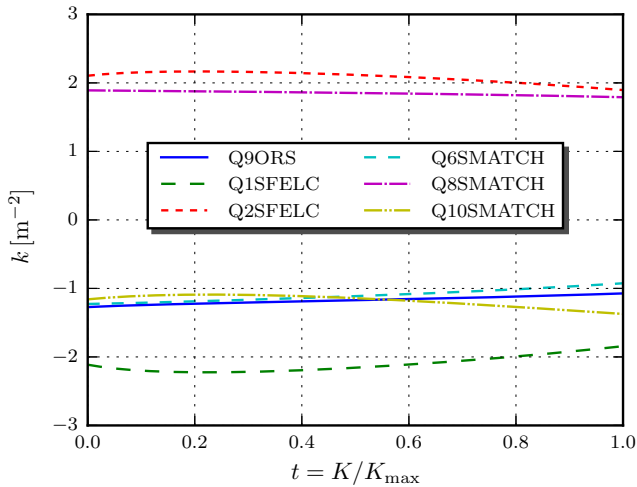}
    \caption{\label{fig:correction}Example of a correction function calculated using the presented method.
Here, the strengths $k$ of the six named quadrupole magnets are used as correction parameters to
compensate for the optics disturbance caused by the sFLASH undulator being closed from $K = 0$ to $K = K_\textrm{max} = 2.63$
at a beam energy of $\SI{693}{\mega\electronvolt}$.
}
\end{figure}

Our implementation utilizes the computer algebra system \texttt{Mathematica} \cite{WolframResearch2014},
which allows for easily calculating the required symbolic transfer matrix
according to equation \eqref{eq:matmult}, as well as setting up the differential equations \eqref{eq:diffeq},
the numerical solution of which is then found by means of a basic Runge-Kutta algorithm.
There are $14$ quadrupole magnets in the vicinity of the sFLASH undulator (see \figref{fig:magnets})
that are included in the \texttt{Mathematica} model and are available as correctors.
As our method produces correction functions for exactly $6$ parameters
there are
\[\binom{14}{6} = 3003 \]
combinations of corrector magnets to choose from.
Investigation shows that a lot of these combinations are prone to produce
solutions featuring impracticably high changes in the quadrupole strengths
or solutions containing a singularity, as mentioned above.
However, we were able to identify a set of known good corrector combinations
for optics similar to the standard optics used in nominal SASE operation.
An example of the correction functions that are obtained by this method is shown in \figref{fig:correction}.

Surely, for any given manipulation, correction functions exist which involve more than six corrector magnets.
In that case, the solution of equation \eqref{Bconst} is no longer unique because, as we have seen,
solutions can be constructed using any combination of six of the magnets and leave the others constant,
which results in our method being not directly applicable.
However, any number of additional constraints, other than those given by the constancy of the transfer matrix,
can easily be incorporated into our method by appending them to $\vec{B}$ in equation \eqref{Bconst}
and selecting the same number of additional correction parameters.

\begin{figure}[bth]
    \centering
    \includegraphics*[width=\linewidth]{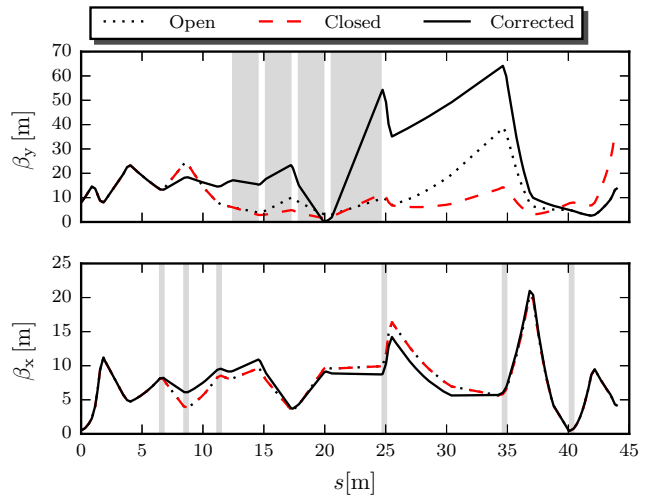}
    \caption{\label{fig:betaplot}Development of the $\beta$-functions in dependence on the longitudinal coordinate $s$.
    The undisturbed state (dotted) is compared to the disturbed state with (solid line) and without
    (dashed) application of the correction.
    Grey areas in the upper plot mark the position of the variable-gap undulator segments.
    In this example, a different set of correction magnets is used.
    Their positions are indicated in the lower plot.
    Note that at the end of the beamline the corrected optical functions
    agree with the \textit{open} case.}
\end{figure}

While our method assures the constancy of the transfer matrix between the first and the last
corrector magnet, it however makes no statement about the behavior of the
optical functions within this correction section.
Using the particle tracking code \texttt{ELEGANT}, the optics can be checked for
unwanted features and the efficacy of the correction can be verified, see \figref{fig:betaplot}.

\section{Measurements}

Diagnostics in the FLASH1 main undulator area include wire scanners at the ends
of the undulator modules. With the help of these devices the beam diameter
can be determined along the undulator.
The relationship between beam size $\sigma$ and $\beta$-function in the vertical
and horizontal direction is given by the well known equation
\begin{equation}
\sigma_{u}(z) = \sqrt{\varepsilon_u\,\beta_u(z)},\quad u \in \{x,y\},
\end{equation}
where $z$ is the longitudinal position and $\varepsilon_u$ denotes the beam emittance.
A successful correction can therefore be verified experimentally, by comparing
the beam sizes along the undulator.

\begin{figure}[bth]
	\centering
	\includegraphics*[width=\linewidth]{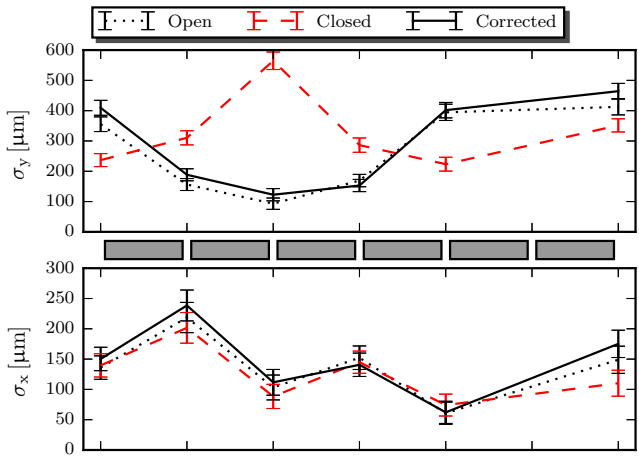}
	\caption{\label{fig:wirescans}Transverse beam sizes measured using the wire scanners between the six FLASH main
	undulator segments (grey boxes in the middle). The \textit{dotted} curve shows the beam sizes at theory optics without any disturbance.
	Measurements of the disturbed case, where the seeding undulator is closed, are depicted by the \textit{dashed} curve.
	The \textit{solid} curve shows beam sizes with corrections applied.
	At the time of measurement, the wire scanner between the last two undulator segments was unavailable \cite{[{Figures \ref{fig:correction}, \ref{fig:betaplot} and \ref{fig:wirescans} were created using \texttt{matplotlib}. }]Hunter2007}.
	}
\end{figure}

The measurements presented in \figref{fig:wirescans} have been conducted with a 1.0-GeV-electron
beam, while the disturbance was introduced by closing all four seeding undulator segments
to their lowest gap value.

As expected, closing the seeding undulator primarily affects the beam size in the vertical direction.
After the optics correction is applied, the beam sizes correspond to those
measured with an open undulator within the range of measuring accuracy.
Thereby, it is shown experimentally that our method is able to restore the initial
course of the optical functions in the main undulator with satisfactory precision.

As mentioned above, the motivation leading to the development of the presented method was
to allow simultaneous operation of both the sFLASH and FLASH1 main undulator.
In recent efforts to establish procedures for three beamline lasing (FLASH1, FLASH2, sFLASH)
the described method is in frequent use \cite{Plath2016}.
Once FLASH1 SASE operation is
established, the sFLASH variable-gap undulator segments are closed to their
target gap values, causing the SASE signal of the main undulator to vanish.
Subsequently, the calculated correction is applied to the chosen corrector quadrupole magnets.
This on-line procedure was able to recover 70\% of the initial FEL photon pulse energy
with only a few orbit corrections applied.

\section{Conclusion and Outlook}
We developed a novel approach to compensate for variations in particle beam optics
that would otherwise disturb the optics in other parts of the machine, if let uncorrected.
It produces continuous, numerical functions for six machine parameters,
compensating the unwanted effect of an arbitrary number of disturbance parameters.
Because of the continuity of the correction function, it is suitable
to be applied simultaneously to the disturbance.
Effectively, this method allows to freely modify the optics locally within a beamline section,
without altering the optics in the rest of the machine.

The method has been implemented at FLASH, with the objective of compensating
the influence of the variable-gap undulator used by the sFLASH experiment
and is in frequent use.
We showed measurements confirming the precise restoration of the initial optics.

A next experimental step will be to apply the correction steadily as the
disturbance arises, which in the end enables us to close the seeding undulator and start
seeding experiments without any notable changes of electron beam quality in the FLASH1 main
undulator.
This compensation constitutes an important milestone for the planned simultaneous operation
of a seeding experiment at FLASH1.
The method will generally be useful for two FELs operated independently of each other in cascaded mode.
Ultimately, it will be beneficial for all optics manipulations that need to stay restricted to a local
section of a long beamline.

\begin{acknowledgments}
We would like to thank all colleagues participating in FLASH operation.
This work has been supported by Federal Ministry
of Education and Research of Germany under contract
No. 05K1GU4 and the German Research
Foundation programme graduate school 1355.
\end{acknowledgments}

\bibliography{bibliography}

\begin{thebibliography}{13}%
\makeatletter
\providecommand \@ifxundefined [1]{%
 \@ifx{#1\undefined}
}%
\providecommand \@ifnum [1]{%
 \ifnum #1\expandafter \@firstoftwo
 \else \expandafter \@secondoftwo
 \fi
}%
\providecommand \@ifx [1]{%
 \ifx #1\expandafter \@firstoftwo
 \else \expandafter \@secondoftwo
 \fi
}%
\providecommand \natexlab [1]{#1}%
\providecommand \enquote  [1]{``#1''}%
\providecommand \bibnamefont  [1]{#1}%
\providecommand \bibfnamefont [1]{#1}%
\providecommand \citenamefont [1]{#1}%
\providecommand \href@noop [0]{\@secondoftwo}%
\providecommand \href [0]{\begingroup \@sanitize@url \@href}%
\providecommand \@href[1]{\@@startlink{#1}\@@href}%
\providecommand \@@href[1]{\endgroup#1\@@endlink}%
\providecommand \@sanitize@url [0]{\catcode `\\12\catcode `\$12\catcode
  `\&12\catcode `\#12\catcode `\^12\catcode `\_12\catcode `\%12\relax}%
\providecommand \@@startlink[1]{}%
\providecommand \@@endlink[0]{}%
\providecommand \url  [0]{\begingroup\@sanitize@url \@url }%
\providecommand \@url [1]{\endgroup\@href {#1}{\urlprefix }}%
\providecommand \urlprefix  [0]{URL }%
\providecommand \Eprint [0]{\href }%
\providecommand \doibase [0]{http://dx.doi.org/}%
\providecommand \selectlanguage [0]{\@gobble}%
\providecommand \bibinfo  [0]{\@secondoftwo}%
\providecommand \bibfield  [0]{\@secondoftwo}%
\providecommand \translation [1]{[#1]}%
\providecommand \BibitemOpen [0]{}%
\providecommand \bibitemStop [0]{}%
\providecommand \bibitemNoStop [0]{.\EOS\space}%
\providecommand \EOS [0]{\spacefactor3000\relax}%
\providecommand \BibitemShut  [1]{\csname bibitem#1\endcsname}%
\let\auto@bib@innerbib\@empty
\bibitem [{\citenamefont {Borland}(2000)}]{Borland2000}%
  \BibitemOpen
  \bibfield  {author} {\bibinfo {author} {\bibfnamefont {M.}~\bibnamefont
  {Borland}},\ }\href@noop {} {\enquote {\bibinfo {title} {{ELEGANT: A flexible
  SDDS-compliant code for accelerator simulation}},}\ } (\bibinfo {year}
  {2000})\BibitemShut {NoStop}%
\bibitem [{\citenamefont {Grote}\ \emph {et~al.}(2016)\citenamefont {Grote}
  \emph {et~al.}}]{MADX}%
  \BibitemOpen
  \bibfield  {author} {\bibinfo {author} {\bibfnamefont {H.}~\bibnamefont
  {Grote}} \emph {et~al.},\ }\href@noop {} {\emph {\bibinfo {title} {The MAD-X
  Program (Methodical Accelerator Design)}}},\ \bibinfo {organization} {CERN},\
  \bibinfo {address} {Geneva} (\bibinfo {year} {2016})\BibitemShut {NoStop}%
\bibitem [{\citenamefont {Krantz}\ and\ \citenamefont
  {Parks}(2002)}]{Krantz2002}%
  \BibitemOpen
  \bibfield  {author} {\bibinfo {author} {\bibfnamefont {S.}~\bibnamefont
  {Krantz}}\ and\ \bibinfo {author} {\bibfnamefont {H.}~\bibnamefont {Parks}},\
  }\href@noop {} {\emph {\bibinfo {title} {The Implicit Function Theorem:
  History, Theory, and Applications}}}\ (\bibinfo  {publisher} {Birkhäuser},\
  \bibinfo {year} {2002})\BibitemShut {NoStop}%
\bibitem [{\citenamefont {Bödewadt}\ \emph {et~al.}(2015)\citenamefont
  {Bödewadt} \emph {et~al.}}]{Boedewadt2015}%
  \BibitemOpen
  \bibfield  {author} {\bibinfo {author} {\bibfnamefont {J.}~\bibnamefont
  {Bödewadt}} \emph {et~al.},\ }in\ \href@noop {} {\emph {\bibinfo {booktitle}
  {Proceedings of IPAC2015}}},\ \bibinfo {series and number} {\bibinfo {number}
  {TUPC3}}\ (\bibinfo {year} {2015})\BibitemShut {NoStop}%
\bibitem [{\citenamefont {Courant}\ and\ \citenamefont
  {Snyder}(1958)}]{Courant1958}%
  \BibitemOpen
  \bibfield  {author} {\bibinfo {author} {\bibfnamefont {E.}~\bibnamefont
  {Courant}}\ and\ \bibinfo {author} {\bibfnamefont {H.}~\bibnamefont
  {Snyder}},\ }\href@noop {} {\bibfield  {journal} {\bibinfo  {journal} {Annals
  of Physics}\ }\textbf {\bibinfo {volume} {3}},\ \bibinfo {pages} {1 }
  (\bibinfo {year} {1958})}\BibitemShut {NoStop}%
\bibitem [{\citenamefont {Gilmore}(2005)}]{Gilmore2005}%
  \BibitemOpen
  \bibfield  {author} {\bibinfo {author} {\bibfnamefont {R.}~\bibnamefont
  {Gilmore}},\ }\href@noop {} {\emph {\bibinfo {title} {Lie Groups, Lie
  Algebras, and Some of Their Applications}}},\ Dover Books on Mathematics\
  (\bibinfo  {publisher} {Dover Publications},\ \bibinfo {year}
  {2005})\BibitemShut {NoStop}%
\bibitem [{\citenamefont {Allgower}\ and\ \citenamefont
  {Georg}(2003)}]{Allgower2003}%
  \BibitemOpen
  \bibfield  {author} {\bibinfo {author} {\bibfnamefont {E.}~\bibnamefont
  {Allgower}}\ and\ \bibinfo {author} {\bibfnamefont {K.}~\bibnamefont
  {Georg}},\ }\href@noop {} {\emph {\bibinfo {title} {Introduction to Numerical
  Continuation Methods}}},\ Classics in Applied Mathematics\ (\bibinfo
  {publisher} {Society for Industrial and Applied Mathematics},\ \bibinfo
  {year} {2003})\BibitemShut {NoStop}%
\bibitem [{\citenamefont {Honkavaara}\ \emph {et~al.}(2014)\citenamefont
  {Honkavaara} \emph {et~al.}}]{Honkavaara2014}%
  \BibitemOpen
  \bibfield  {author} {\bibinfo {author} {\bibfnamefont {K.}~\bibnamefont
  {Honkavaara}} \emph {et~al.},\ }in\ \href@noop {} {\emph {\bibinfo
  {booktitle} {Proceedings of FEL2014}}},\ \bibinfo {series and number}
  {\bibinfo {number} {WEB05}}\ (\bibinfo {year} {2014})\BibitemShut {NoStop}%
\bibitem [{\citenamefont {Quattromini}\ \emph {et~al.}(2012)\citenamefont
  {Quattromini} \emph {et~al.}}]{Quattromini2012}%
  \BibitemOpen
  \bibfield  {author} {\bibinfo {author} {\bibfnamefont {M.}~\bibnamefont
  {Quattromini}} \emph {et~al.},\ }\href@noop {} {\bibfield  {journal}
  {\bibinfo  {journal} {Phys. Rev. ST Accel. Beams}\ }\textbf {\bibinfo
  {volume} {15}},\ \bibinfo {pages} {080704} (\bibinfo {year}
  {2012})}\BibitemShut {NoStop}%
\bibitem [{\citenamefont {Tischer}\ \emph {et~al.}(2010)\citenamefont {Tischer}
  \emph {et~al.}}]{Tischer2010}%
  \BibitemOpen
  \bibfield  {author} {\bibinfo {author} {\bibfnamefont {M.}~\bibnamefont
  {Tischer}} \emph {et~al.},\ }in\ \href@noop {} {\emph {\bibinfo {booktitle}
  {Proceedings IPAC2010}}},\ \bibinfo {series and number} {\bibinfo {number}
  {WEPD014}}\ (\bibinfo {year} {2010})\BibitemShut {NoStop}%
\bibitem [{\citenamefont {{Wolfram Research,
  Inc.}}(2014)}]{WolframResearch2014}%
  \BibitemOpen
  \bibfield  {author} {\bibinfo {author} {\bibnamefont {{Wolfram Research,
  Inc.}}},\ }\href@noop {} {\enquote {\bibinfo {title} {Mathematica},}\ }
  (\bibinfo {year} {2014})\BibitemShut {NoStop}%
\bibitem [{\citenamefont {Hunter}(2007)}]{Hunter2007}%
  \BibitemOpen
  \bibfield  {author} {\bibinfo {author} {\bibfnamefont {J.~D.}\ \bibnamefont
  {Hunter}},\ }\href@noop {} {\bibfield  {journal} {\bibinfo  {journal}
  {Computing In Science \& Engineering}\ }\textbf {\bibinfo {volume} {9}},\
  \bibinfo {pages} {90} (\bibinfo {year} {2007})}\BibitemShut {NoStop}%
\bibitem [{\citenamefont {Plath}\ \emph {et~al.}(2016)\citenamefont {Plath}
  \emph {et~al.}}]{Plath2016}%
  \BibitemOpen
  \bibfield  {author} {\bibinfo {author} {\bibfnamefont {T.}~\bibnamefont
  {Plath}} \emph {et~al.},\ }\href@noop {} {\bibfield  {journal} {\bibinfo
  {journal} {TBP}\ } (\bibinfo {year} {2016})}\BibitemShut {NoStop}%
\end{thebibliography}%

\end{document}